\documentclass[12pt]{article}

\begin{document}


\def\a{\alpha}
\def\b{\beta}
\def\c{\varepsilon}
\def\d{\delta}
\def\e{\epsilon}
\def\f{\phi}
\def\g{\gamma}
\def\h{\theta}
\def\k{\kappa}
\def\l{\lambda}
\def\m{\mu}
\def\n{\nu}
\def\p{\psi}
\def\q{\partial}
\def\r{\rho}
\def\s{\sigma}
\def\t{\tau}
\def\u{\upsilon}
\def\v{\varphi}
\def\w{\omega}
\def\x{\xi}
\def\y{\eta}
\def\z{\zeta}
\def\D{{\mit \Delta}}
\def\G{\Gamma}
\def\H{\Theta}
\def\L{\Lambda}
\def\F{\Phi}
\def\P{\Psi}
\def\S{\Sigma}
\def\V{\varPsi}

\def\o{\over}
\newcommand{\sla}[1]{#1 \llap{\, /}}

\newcommand{\beq}{\begin{eqnarray}}
\newcommand{\eeq}{\end{eqnarray}}
\newcommand{\gsim}{ \mathop{}_{\textstyle \sim}^{\textstyle >} }
\newcommand{\lsim}{ \mathop{}_{\textstyle \sim}^{\textstyle <} }
\newcommand{\vev}[1]{ \left\langle {#1} \right\rangle }
\newcommand{\bra}[1]{ \langle {#1} | }
\newcommand{\ket}[1]{ | {#1} \rangle }

\newcommand{\EV}{ \,{\rm eV} }
\newcommand{\KEV}{ \,{\rm keV} }
\newcommand{\MEV}{ \,{\rm MeV} }
\newcommand{\GEV}{ \,{\rm GeV} }
\newcommand{\TEV}{ \,{\rm TeV} }


\baselineskip 0.7cm

\begin{titlepage}

\begin{flushright}
UT-05-01 \\
\end{flushright}

\vskip 1.35cm
\begin{center}
{\large \bf
Strongly Coupled Gauge Mediation
}
\vskip 1.2cm
Izawa K.-I.~and T.~Yanagida
\vskip 0.4cm

{\it Department of Physics, University of Tokyo,\\
     Tokyo 113-0033, Japan}

\vskip 1.5cm

\abstract{
A model of gauge mediation with $m_{3/2}\simeq {\cal O}(1)\EV$
is considered to motivate $100\TEV$ colliders.
Massive mediators with standard-model and supersymmetry-breaking
gauge quantum numbers let low-scale dynamics induce sizable
soft masses of the standard-model superpartners.
We circumvent potential phenomenological difficulties
that such low-scale models tend to cause. 
}
\end{center}
\end{titlepage}

\setcounter{page}{2}

\section{Introduction}

Light gravitino of mass $\lsim 10\EV$ is very interesting, since it does
not have any astrophysical and cosmological problems
\cite{gravitino}.
Furthermore, it
brings an intriguing possibility of the axion dark matter
\cite{axion}
back into our attention. This leads us to construct an
explicit model of gauge mediation
\cite{mediation}
with the gravitino mass $m_{3/2}\simeq {\cal O}(1)\EV$.%
\footnote{Several attempts have been made
to construct such models of low-scale gauge mediation.
See Ref.\cite{Iza,Nom}.
}

\section{SUSY Breaking}

We adopt an $Sp(2)$
gauge theory with $6$ chiral superfields $Q_i$
in the fundamental $4$-dimensional representation,
where $i$ is a flavor index ($i=1,\cdots,6$)
and the gauge index is omitted.
Without a superpotential, this theory has a flavor $SU(6)_F$ symmetry.
This $SU(6)_F$ symmetry
is explicitly broken down to a flavor $Sp(3)_F$
by a superpotential in our model.
That is, we add gauge singlets $Z^a$ ($a=1,\cdots,14$)
to obtain the tree-level superpotential
\begin{eqnarray}
 W_0 = Z^a (QQ)_a,
\end{eqnarray}
where $(QQ)_a$ denotes
a flavor $14$-plet
of $Sp(3)_F$ given by a suitable combination of gauge invariants $Q_iQ_j$.

The effective superpotential
which may describe the dynamics of the $Sp(2)$ gauge interaction
is given by
\begin{eqnarray}
 W_{eff}=X({\rm Pf} V_{ij} - \Lambda^{6}) + Z^a V_a
\label{dynamical_potential}
\end{eqnarray}
in terms of low-energy degrees of freedom
\begin{eqnarray}
 V_{ij} \sim Q_iQ_j, \quad V_a \sim (QQ)_a,
\end{eqnarray}
where $X$ is an additional chiral superfield and
$\L$ denotes a dynamical scale of the gauge interaction.
This effective superpotential implies that,
among the gauge invariants $Q_iQ_j$, the $Sp(3)_F$ singlet $(QQ)$
condenses as
%
$\langle (QQ) \rangle = \Lambda^2$.
%

Let us consider
\beq
 W=W_0+\l Z(QQ).
\eeq
When the coupling $\lambda$ is small, the effective superpotential
$W_{eff}$ implies that we obtain the following vacuum expectation
values:
\begin{equation}
  \label{vevs}
  \langle (QQ) \rangle \simeq \Lambda^2, \quad \langle V_a \rangle \simeq 
  0.
\end{equation}
Then the low-energy effective superpotential may be approximated by
\begin{equation}
  \label{effPot}
  W_{eff} \simeq \lambda \Lambda^2 Z,
\end{equation}
which finally yields dynamical supersymmetry (SUSY) breaking
\cite{Yan}
\beq
 F_Z \simeq \lambda \Lambda^2.
\eeq

There seem two alternative possibilities in this SUSY-breaking dynamics:
$\vev{\l Z}=0$ or $\vev{\l Z} \neq 0$
\cite{Hot,Cha}.
We first bear in mind the latter case $\vev{\l Z} \neq 0$
and later in section 4, we proceed to the other case $\vev{\l Z}=0$.

\section{Massive Mediators}

We introduce mediator chiral superfields $\F, {\bar \F}$,
which are (anti-)fundamentals under the standard-model (SM)
gauge group%
\footnote{Namely, for definiteness,
we assume that $\F$ transforms as the anti-down quark
and the lepton doublet.
Since these mediators have the SM quantum numbers,
they can decay into the SM particles
(through higher-dimensional interactions) without yielding
dangerously long-lived cosmological relics.
The singlets $Z^a$ and the bound states $(QQ)_a$
can also decay into the SM particles if one allows
flavor $Sp(3)_F$ violating interactions such as $Z(QQ)_a$.}
and fundamentals
under the SUSY-breaking $Sp(2)$ gauge group.
The superpotential of our model for the mediators
is given by
\cite{Iza}
\begin{equation}
 \label{REQ}
 W_{med} = m \F {\bar \F},
\end{equation}
where $m (\gsim 4\pi \Lambda)$ is a mass parameter.
This is a supersymmetric mass term similar to that of
the Higgs doublets (so-called $\mu$-term)
in the supersymmetric SM.
Notice that the $Sp(2)$ theory is asymptotically free
for a pair of the chiral multiplets $\F, {\bar \F}$. 

The mediators $\F$ and ${\bar \F}$,
which are massive,
may be integrated out.%
\footnote{Through an effective identification
$X \sim m \F {\bar \F}/\L^6$,
we see
\beq
 \vev{m \F {\bar \F}} \simeq {5 \o 3}\vev{\l Z}\L^2,
 \nonumber
\eeq
where the factor 5 comes from the mediator number of the
$Sp(2)$ quartet pairs. 
We suppose that this condensation is a SM singlet
contained in $\F \otimes {\bar \F}$ and respects the SM gauge symmetry.}
Owing to their SM and SUSY-breaking gauge interactions,
they induce SUSY-breaking soft masses of the SM superpartners.
When the $Sp(2)$ gauge interaction becomes strong
around the mass scale $m$ with the mediators decoupling
\cite{Iza},
we obtain
\beq
 m_0^2 \simeq 4\left({\a \o 4\pi}\right)^2{1 \o m^2}|\l F_Z|^2
 \simeq 4\left({\a \o 4\pi}\right)^2{|\l^2 \L^2|^2 \o m^2},
\eeq
and
\beq
 m_{1/2} \simeq 4{\a \o 4\pi}{1 \o m^2}{\l F_Z \vev{\l Z}^*}
 \simeq 4{\a \o 4\pi}{\l^2 \L^2 \o m^2}{\vev{\l Z}^*},
\eeq
where the overall factor $4$ takes into account
the mediator number of SM flavors
and naive dimensional counting
\cite{Lut}
is applied.
Here, $\a$ corresponds to the relevant SM gauge couplings
to the sfermions and the gauginos.
On the other hand, the gravitino mass is obtained as
\beq
 m_{3/2} \simeq {F_Z \o \sqrt{3}M_{Pl}}
 \simeq {\l \L^2 \o \sqrt{3}M_{Pl}},
\eeq
where $M_{Pl}$ denotes the reduced Planck scale:
$M_{Pl} \simeq 2.4 \times 10^{18}\GEV$.

For $m \simeq 4\pi \L$,
\beq
 m_0^2
 \simeq \left({\a \o 4\pi}\right)^2{1 \o 4\pi^2}|\l^2 \L|^2,
\eeq
and
\beq
 m_{1/2}
 \simeq {\a \o 4\pi}{1 \o \pi}{\l^2 \L},
\eeq
where we have used $\vev{\l Z} \simeq 4\pi \L$.
We see that $m_{3/2} \simeq {\cal O}(1)\EV$ is achieved
for $\sqrt{\l} \L \simeq 100\TEV$
with $m_0 \simeq m_{1/2} \simeq {\cal O}(100)\GEV$.

\section{R-axion and $\mu$-term}

The above setup respects R-symmetry and $\vev{\l Z} \simeq 4\pi \L$
results in an R-axion
\cite{Bag},
which is marginally consistent to astrophysical and
cosmological constraints
for $\sqrt{\l} \L$ as low as $100\TEV$.
We note here that we can safely satisfy these constraints
as follows.
We further introduce additional chiral superfields $\psi, {\bar \psi}$,
which are (anti-)fundamentals under the SM
gauge group and singlets
under the SUSY-breaking $Sp(2)$ gauge group.
The superpotential for these fields
is given by
\begin{equation}
 \label{REQ}
 W_{add} = (M + kZ) \psi {\bar \psi},
\end{equation}
where $M (\gg |\vev{k Z}|)$ is an additional mass parameter
and $k$ a coupling constant.
This interaction makes the SUSY-breaking vacuum $|\vev{k Z}| \ll M$
a local one with the supersymmetric minimum far away at $kZ=-M$.%
\footnote{Another way to avoid an unwanted R-axion
(and a supersymmetric minimum) is to introduce a singlet $Y$
with superpotential terms such as $(QQ)Y+Y^3$
\cite{Hot}.}

The above terms explicitly break the R symmetry and
make the would-be R-axion completely harmless for $M \ll M_{Pl}$.%
\footnote{They also break an accidental $Z_3$ symmetry
in the SUSY-breaking dynamics and eliminate the corresponding
domain wall problem. The scale $M$ can be as large as
an intermediate scale ${\cal O}(10^{14})\GEV$
for $\sqrt{\l} \L \simeq 100\TEV$
to make the would-be R-axion sufficiently massive.}
Note that we also obtain additional contributions
\beq
 \D m_0^2 \simeq \left({\a \o 4\pi}\right)^2{|kF_Z|^2 \o M^2},
\eeq
and
\beq
 \D m_{1/2} \simeq {\a \o 4\pi}{kF_Z \o M}.
\eeq
This contribution can make gauginos massive even when $\vev{\l Z}=0$.

For the case $\vev{\l Z}=0$,
the strongly coupled mediators provide a straightforward
origin of the supersymmetric Higgsino mass term.
Let us consider a superpotential of the form
\beq
 W=h Q {\bar \F} H + {\bar h} Q \F {\bar H},
\eeq
where $h$ and ${\bar h}$ denote coupling constants.
Then, the integration of heavy modes induces the $\mu$-term
\beq
 W_{eff} \simeq -h {\bar h}{\vev{(QQ)} \o m}H{\bar H},
\eeq
which yields
\beq
 \mu \simeq -h {\bar h}{\L^2 \o m},
\eeq
without a large B-term.%
\footnote{The case $\vev{\l Z} \neq 0$ suffers from a large B-term
in the present setup.}
It is interesting that $\mu$ and $m_0$ obey the same scaling law
as $\L^2/m$.

\section{Conclusion}

We confirm
the presence of an explicit model for the light gravitino of mass
$m_{3/2}\simeq {\cal O}(1)\EV$ without any phenomenological difficulties.
That constitutes one of the motivations for
experiments at barely imagined colliders of, say, $100\TEV$.
We also note that the present setting is indeed applicable
for a wide range of the gravitino mass $m_{3/2} \simeq 1\EV-100\TEV$,
among which only experiments tell what is realized in Nature.


\end{document}